\documentclass{midl} 

\usepackage{mwe} 
\usepackage{float}
\usepackage{hyperref}
\jmlrproceedings{}{}
\jmlrpages{}
\jmlryear{}

\title[Automatic segmentation of spinal MS lesions: How to generalize across MRI contrasts?]{Automatic segmentation of spinal multiple sclerosis lesions: How to generalize across MRI contrasts?}

\midlauthor{\Name{Olivier Vincent\nametag{$^{1}$}} \Email{ovincent.poly@gmail.com}\\
\Name{Charley Gros\nametag{$^{1}$}} \Email{charley.gros@polymtl.ca}\\
\Name{Joseph Paul Cohen\nametag{$^{2}$}} \Email{joseph@josephpcohen.com}\\
\Name{Julien Cohen-Adad\nametag{$^{1,3}$}} \Email{jcohen@polymtl.ca}\\
\addr $^{1}$ NeuroPoly Lab, Institute of Biomedical Engineering, Polytechnique Montreal, Canada 
\\
\addr $^{2}$ Mila, University of Montreal, Canada \\
\addr $^{3}$ Functional Neuroimaging Unit, CRIUGM, University of Montreal, Canada 
}

\usepackage{parskip}

\begin{document}

\maketitle

\begin{abstract}

 Despite recent improvements in medical image segmentation, the ability to generalize across imaging contrasts remains an open issue. To tackle this challenge, we implement Feature-wise Linear Modulation (FiLM) to leverage physics knowledge within the segmentation model and learn the characteristics of each contrast. Interestingly, a well-optimised U-Net reached the same performance as our FiLMed-Unet on a multi-contrast dataset (0.72 of Dice score), which suggests that there is a bottleneck in spinal MS lesion segmentation different from the generalization across varying contrasts. This bottleneck likely stems from inter-rater variability, which is estimated at 0.61 of Dice score in our dataset.
\end{abstract}

\begin{keywords}
    Deep Learning, Segmentation, MRI, Spinal cord, Multiple Sclerosis
\end{keywords}

\section{Introduction}

Multiple Sclerosis (MS) is the most prevalent autoimmune disease of the central nervous system \cite{Berer2014-ww}. Lesion quantification on both brain and spinal cord MRI data is part of the diagnosis criteria for MS, \cite{Thompson2018-ji} and has been extensively used in clinical studies \cite{Kearney2015-xj}. Although recent methods based on deep convolutional neural networks (CNNs) showed promising results \cite{Gros2018-ss,McCoy2019-ce}, they are hampered by major issues \cite{LITJENS201760}. One of the issues is the inability of CNN models to generalize to heterogeneous imaging parameters (e.g. MR field strength or manufacturer, image contrast, resolution and field of view) that were not represented in the training data.

\section{Methods}

To address the generalization problem, we adapted the Feature-wise Linear Modulation (FiLM) \cite{Perez2017-od, dumoulin2016learned} approach to the segmentation task. FiLM enables us to modulate CNNs features based on non-image metadata as illustrated in \figureref{fig:film}. In order to facilitate the model generalization, we input the MRI contrast (e.g. T2-weighted) in the FiLM generator, instead of directly inputting MR acquisition parameters which, based on preliminary investigations, would produce too many degrees of freedom and non-linearity issues across parameters. Each FiLM generator (multi-layer perceptron) optimises $\gamma,\beta$ based on the contrast information ($\boldsymbol{z}$). Each U-Net feature map ($\boldsymbol{x}$) is then linearly-modulated by these FiLM parameters, such that :
\begin{align*}
    \text{FiLM}(\boldsymbol{z}) = \gamma( \boldsymbol{z}) \circ  \boldsymbol{x} + \beta( \boldsymbol{z})    
\end{align*}

\begin{figure}[htb]
\floatconts
  {fig:film}
  {\caption{FiLM architecture using MRI contrast type in input. A FiLM layer is added after each convolutional block of the U-Net to modulate the CNN feature maps. The FiLMedUnet is trained end-to-end.}}
  {\includegraphics[width=1\linewidth]{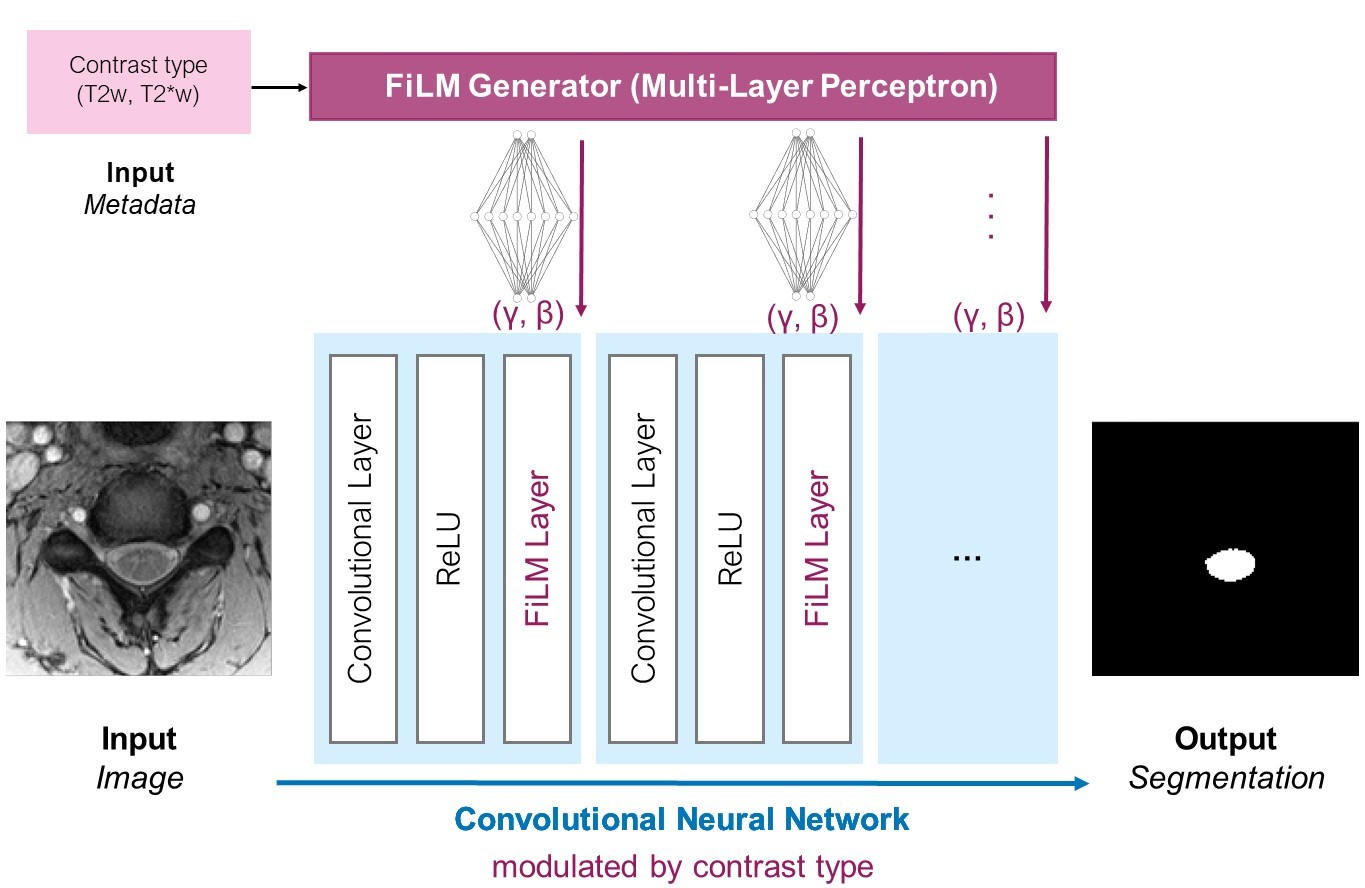}}
\end{figure}

We compared this new approach with a traditional U-Net \cite{Ronneberger2015-sf}, on ‘real-world’ data from $642$ MS patients, acquired by thirteen centers, yielding $2,549$ MR volumes (T2-weighted or T2*-weighted, $38,855$ axial slices in total), spanning a large range of acquisition parameters (e.g. resolution, orientation, field of view). To alleviate the issue of class imbalance, slices were cropped around the region of interest using the spinal cord segmentation (48x48 pixels). Data augmentation was performed on both the MRI data (random affine and elastic transformations) and by altering the ground truth segmentation via a series of morphological and affine realistic operations. This was done to simulate rater uncertainty at the boundary of lesions and have the network learn these uncertainties.

Training was done on axial slices using Dice loss \cite{milletari2016vnet} with the Adam optimizer \cite{kingma2014adam} and a $60/20/20$\% training / validation / testing random split of the dataset. Hyperparameters such as learning rate scheduler, batch size and U-Net depth were optimised using a grid search. Models were implemented in PyTorch 1.2 \cite{Paszke2019-wx} and trained on an NVIDIA P100 GPU, which took ~7h. The implementation is open source and available on Github: \url{https://github.com/neuropoly/ivado-medical-imaging}.

\section{Results}

As shown in \tableref{tab:dice}, models have almost indistinguishable performance when optimised. U-Net yielded negligible performance difference when trained and tested on single-contrast dataset (i.e. T2w or T2*w) compared with multi-contrast dataset (i.e. T2w and T2*w), suggesting that contrast generalization is not a bottleneck, at least in this dataset and for this MS lesion segmentation task. This observation is consistent with the fact that FiLM does not reach higher performance than an optimised U-Net on this dataset.

\begin{table}[htbp]
\floatconts
  {tab:dice}%
  \centering
  {\begin{tabular}{|c|c|}
  \hline
  \bfseries Training configuration & \bfseries Dice score (higher is better)\\
  \hline
  U-Net T2w only & $0.72$\\
  \hline
  U-Net T2*w only & $0.73$\\
  \hline
  U-Net T2w + T2*w & $0.72$\\
  \hline
  FiLMed-Unet T2w + T2*w & $0.72$\\
  \hline
  \end{tabular}}
  {\caption{Results comparison between the U-Net and our FiLMed-U-Net in terms of Dice on the testing dataset, including T2w (top row) or T2*w (second row) or T2w and T2*w (last rows) data. }}%
\end{table}

\section{Conclusion}

In this paper we implemented FiLM to modulate U-Net segmentation based on MRI contrast type. Results show that a simple U-Net can achieve the same performance as FiLM, both on single and multi-contrast datasets. This unfortunate result however highlights a bottleneck in spinal MS lesion automatic segmentation, and likely in medical image segmentation in general: a high inter-rater variability, as also been reported in brain studies \cite{Carass2017-eh}. Inter-rater variability had a Dice of 0.61 in our dataset \cite{Gros2018-ss}, which is lower than our results of 0.72. The difference  could possibly be explained by some overfitting on certain rater styles. Future work will encode the rater identity into the CNN learning in order to account for rater style and expertise. This would enable the model to learn the difference between multiple rating styles, and to choose a desired style at inference time.

 \midlacknowledgments{The authors would like to thank Lucas Rouhier, Anthime Bucquet, Valentine Louis-Lucas and Christian Perone from the IVADO medical imaging team for helpful discussions. Funded by IVADO [EX-2018-4], Canada Research Chair in Quantitative Magnetic Resonance Imaging [950-230815], the Canadian Institute of Health Research [CIHR FDN-143263], the Canada Foundation for Innovation [32454, 34824], the Fonds de Recherche du Québec - Santé [28826], the Fonds de Recherche du Québec - Nature et Technologies [2015-PR-182754], the FRQNT Strategic Clusters Program (2020‐RS4‐265502 ‐ Centre
UNIQUE ‐ Union Neurosciences \& Artificial Intelligence – Quebec, the Natural Sciences and Engineering Research Council of Canada [RGPIN-2019-07244], the Canada First Research Excellence Fund (IVADO and TransMedTech), the Courtois NeuroMod project and the Quebec BioImaging Network [5886, 35450], Spinal Research and Wings for Life (INSPIRED project).}

\bibliography{references}


\end{document}